\documentclass[onecolumn,amsmath,amssymb]{qm2008_abs}
\usepackage{graphicx}
\usepackage{dcolumn}
\usepackage{bm}
\begin{document}
\def\prl{{\em Phys. Rev. Lett. }}
\def\prc{{\em Phys. Rev. {\bf C} }}
\def\prd{{\em Phys. Rev. {\bf D} }}
\def\jap{{\em J. Appl. Phys. }}
\def\ajp{{\em Am. J. Phys. }}
\def\nima{{\em Nucl. Instr. and Meth. Phys. {\bf A} }}
\def\npa{{\em Nucl. Phys. {\bf A}}}
\def\npb{{\em Nucl. Phys. {\bf B}}}
\def\epjc{{\em Eur. Phys. J. {\bf C}}}
\def\epjst{{\em Eur. Phys. J: {\bf Spec. Topics}}}
\def\jpg{{\em J. Phys. G: Nucl. Part. Phy.}}
\def\plb{{\em Phys. Lett. {\bf B}}}
\def\mpla{{\em Mod. Phys. Lett. {\bf A}}}
\def\pr{{\em Phys. Rep.}}
\def\zpc{{\em Z. Phys. {\bf C}}}
\def\zpa{{\em Z. Phys. {\bf A}}}

\title{{\Large Transverse Energy Measurement in Au+Au Collisions by the STAR
Experiment}}

\bigskip
\bigskip
\author{\large Raghunath Sahoo (for the STAR Collaboration)}
\email{Raghunath.Sahoo@subatech.in2p3.fr}
\affiliation{Institute of Physics, Bhubaneswar, India - 751005 and \\
SUBATECH, 4, Rue Alfred Kastler, BP 20722 - 44307 Nantes Cedex 3, France}
\bigskip
\bigskip

\begin{abstract}
\leftskip1.0cm
\rightskip1.0cm
Transverse energy ($E_T$) has been measured with both of its components,
namely hadronic ($E_T^{had}$) and electromagnetic ($E_T^{em}$) in a common
phase space at mid-rapidity for 62.4 GeV Au+Au collisions by the STAR 
experiment. $E_T$ production with centrality and $\sqrt{s_{NN}}$ is
studied with similar measurements from SPS to RHIC and is compared with
a final state gluon saturation model (EKRT). The most striking feature is the 
observation of a nearly constant value of $E_T/N_{ch} \sim 0.8$ GeV from AGS, 
SPS to RHIC. The initial energy density estimated by the boost-invariant Bjorken 
hydrodynamic model, is well above the critical density for a deconfined matter 
of quarks and gluons predicted by lattice QCD calculations. 
\end{abstract}

\maketitle

\section{Introduction}
Extreme conditions of high temperature and densities could be created in 
relativistic heavy ion collisions making a scenario to study the deconfined 
state of quarks and gluons called Quark Gluon Plasma (QGP). The interactions 
can be characterized 
in terms of the global variables like transvesre energy and the charged 
particle multiplicity. These variables are closely related to the collisions 
geometry and are very important in understanding the global properties of 
the system created during heavy ion collisions. $E_T$ is the energy created 
transverse to the beam direction and is generated by the initial scattering 
of the partonic constituents of the incoming nuclei and probably also by the 
produced partons and hadrons \cite{jacob, wang}. $E_T$ measurement gives an 
estimation of the initial Bjorken energy density ($\epsilon_{Bj}$) produced 
in the fireball and also helps in studying the particle production mechanism. 

\section{Data Analysis}
We have analyzed the 62.4 GeV Au+Au minimum-bias STAR data for RHIC run 2004.
The detectors used in this analysis include the Time Projection Chamber (TPC) and
Barrel Electromagnetic Calorimeter (BEMC) in a common phase space 
($0 < \eta < 1$ and full azimuthal coverage). TPC uncorrected mid-rapidity
multiplicity within $|\eta| < 0.5$ and $|V_z| < 30 $ cm, is used for 
the centrality selection. The analysis method adopted here provides an 
independent event-by-event measurement of both the components of transverse 
energy i.e. $E_T^{had}$ and $E_T^{em}$. $E_T^{had}$ is obtained from the 
TPC reconstructed tracks after taking into account the long-lived neutral hadrons 
which could not be detected by the TPC. $E_T^{em}$ is estimated from the energy 
deposited in the calorimeter towers after correcting for the hadronic contaminations, 
by projecting hadronic tracks onto BEMC. The details of the transverse energy 
estimation procedure is discussed in Ref. \cite{starEt, thesis}.

\begin{figure}
\begin{center}
\includegraphics[width=2.3in]{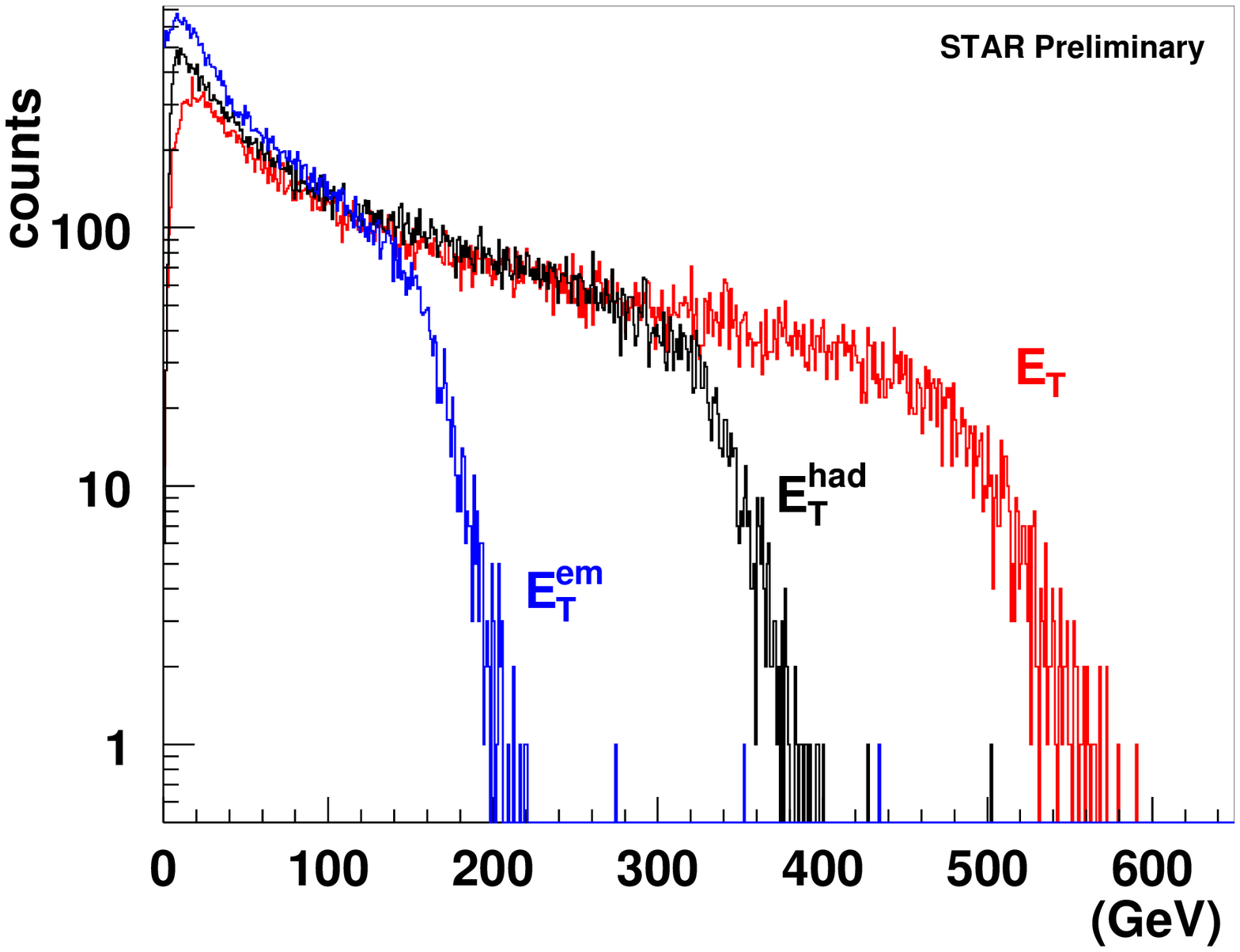}
\includegraphics[width=2.3in]{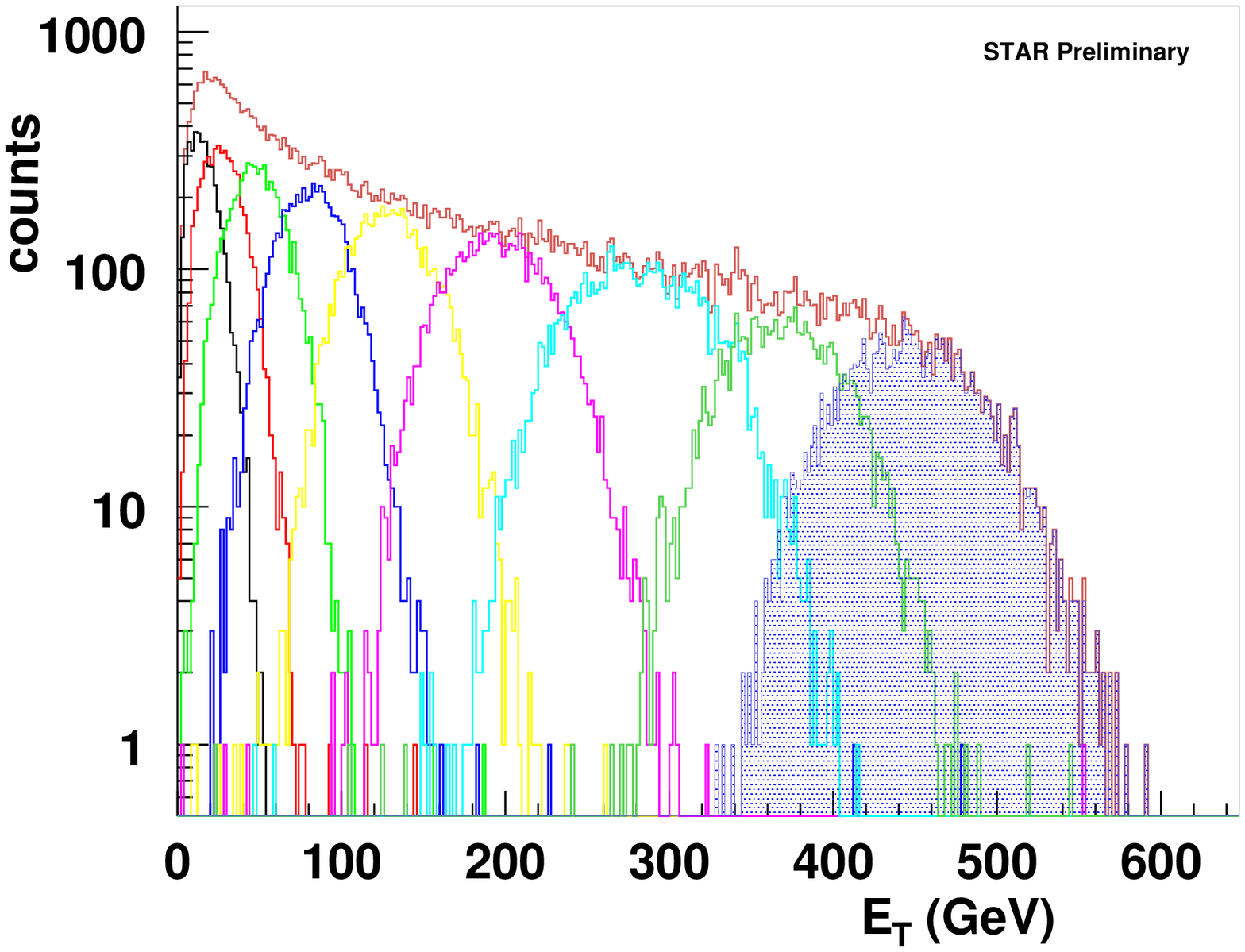}
\caption{The minimum-bias distributions of $E_T$ and its components
(left) and $E_T$ distribution for all centrality classes (right), 
for 62.4 GeV Au+Au collisions.}
\label{etD}
\end{center}
\end{figure}

\section{Results and Discussion}

\begin{figure}
\begin{center}
\includegraphics[width=2.3in]{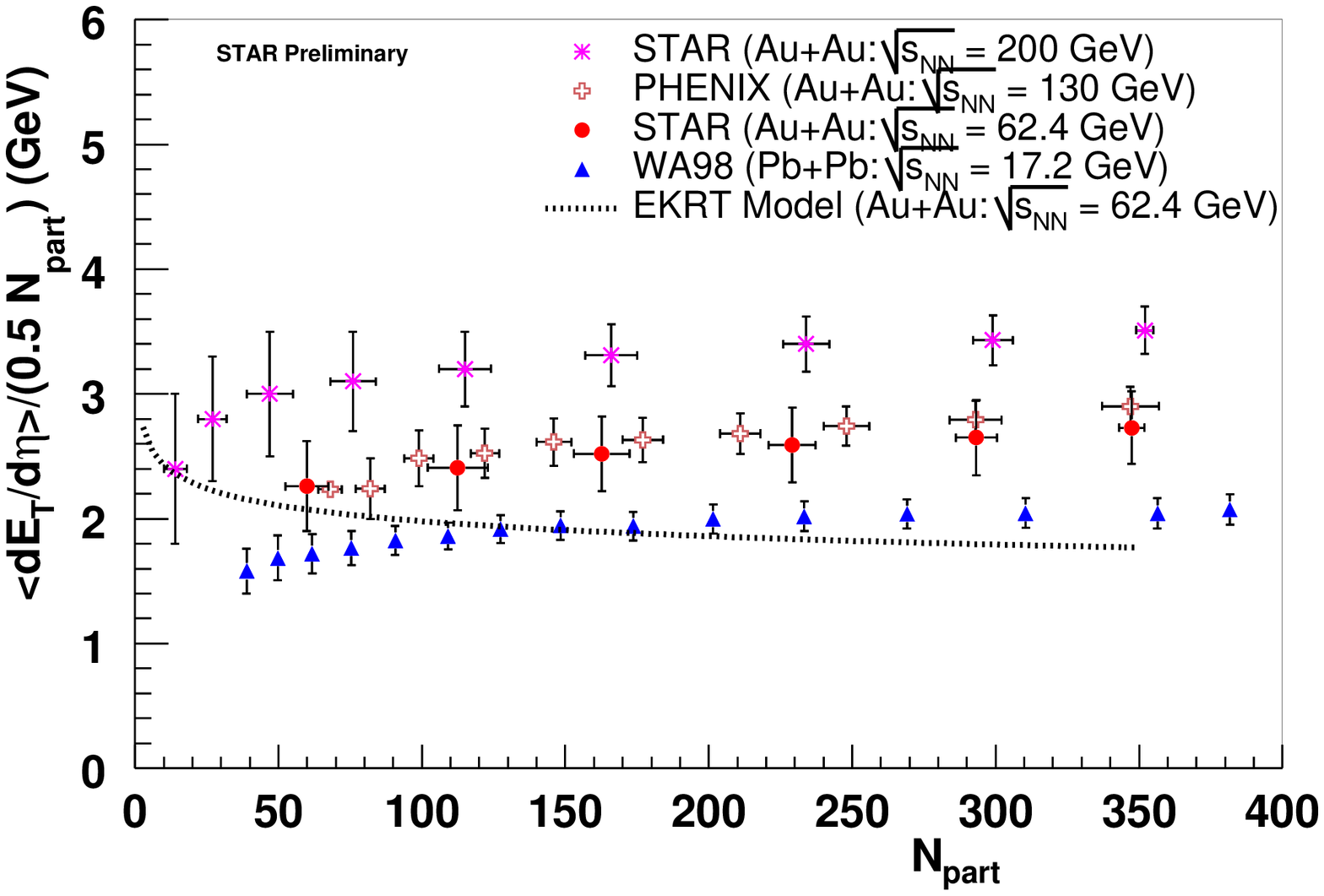}
\includegraphics[width=2.3in]{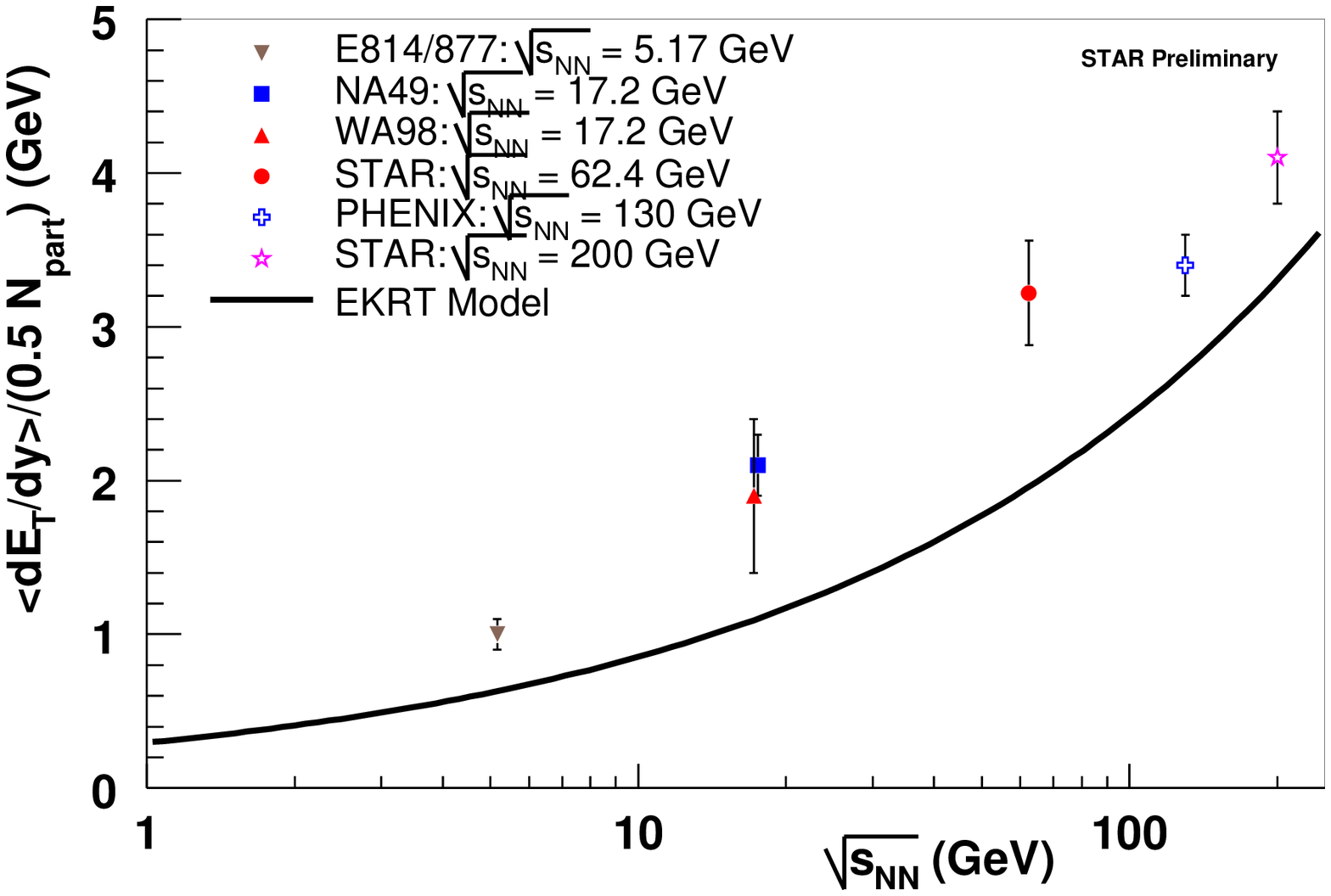}
\caption{The mid-rapidity $<dE_T/d\eta>$ per $N_{part}$ pair vs $N_{part}$
(left) compared to the EKRT model prediction for 62.4 GeV Au+Au data and
$<dE_T/dy>$ per $N_{part}$ pair vs $\sqrt{s_{NN}}$ showing the logarithmic 
growth from AGS-SPS to RHIC (right). The solid line is the EKRT model prection.}
\label{etEKRT}
\end{center}
\end{figure}

 Figure~\ref{etD} (left) shows the event-by-event minimum-bias $E_T$ and its components
with the right figure showing $E_T$ distribution for all centrality 
classes which are gaussian in nature. The minimum-bias distributions show a
peak and a sharp drop-off at low $E_T$ end corresponding to peripheral
collisions. It reaches a broad plateau at the middle which corresponds to
mid-central collisions. This is dominated by nuclear geometry. The higher 
values of $E_T$ correspond to the most central collisions having a ``knee'' 
leading to a fall off which is very steep for large acceptances and less
for small acceptances. For the top $5\%$ central collisions 
$<dE_T/d\eta> = 474 \pm 51 ~GeV$. The variation of $<dE_T/d\eta>$ per 
$N_{part}$ pair as a function of centrality is shown in 
Figure~\ref{etEKRT} (left) for 62.4 GeV Au+Au collisions along with
similar measurements from Pb+Pb 17.3 GeV at SPS \cite{wa98} and Au+Au 130 \cite{phenix}, 
200 GeV \cite{starEt} at RHIC. The corresponding EKRT model \cite{ekrt} prediction 
for 62.4 GeV Au+Au collisions is shown by the dotted line. 
The data show a similar centrality
behavior at different energies, whereas the EKRT model doesn't agree with the data.
For top central collisions, the variation of $<dE_T/dy>$ per $N_{part}$ pair as a function
of $\sqrt{s_{NN}}$ is shown in Figure~\ref{etEKRT} (right) with similar 
measurements at other energies along with the EKRT model
prediction which is shown by the thick line. $<dE_T/dy>/(0.5 N_{part})$ increases 
logarithmically with $\sqrt{s_{NN}}$. However, the EKRT model shows an underestimation 
of data at various energies.

\begin{figure}
\begin{center}
\includegraphics[width=2.3in]{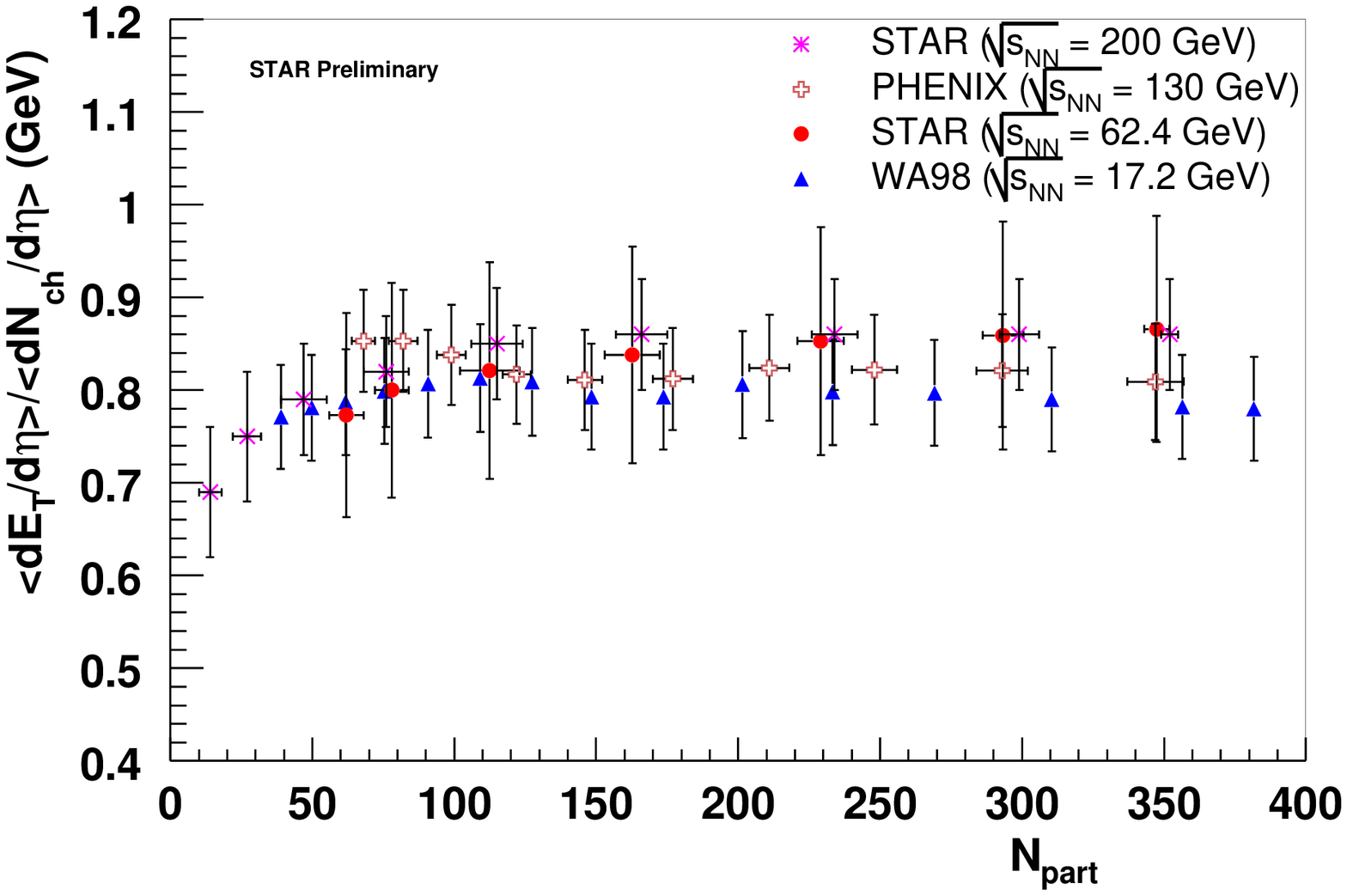}
\includegraphics[width=2.3in]{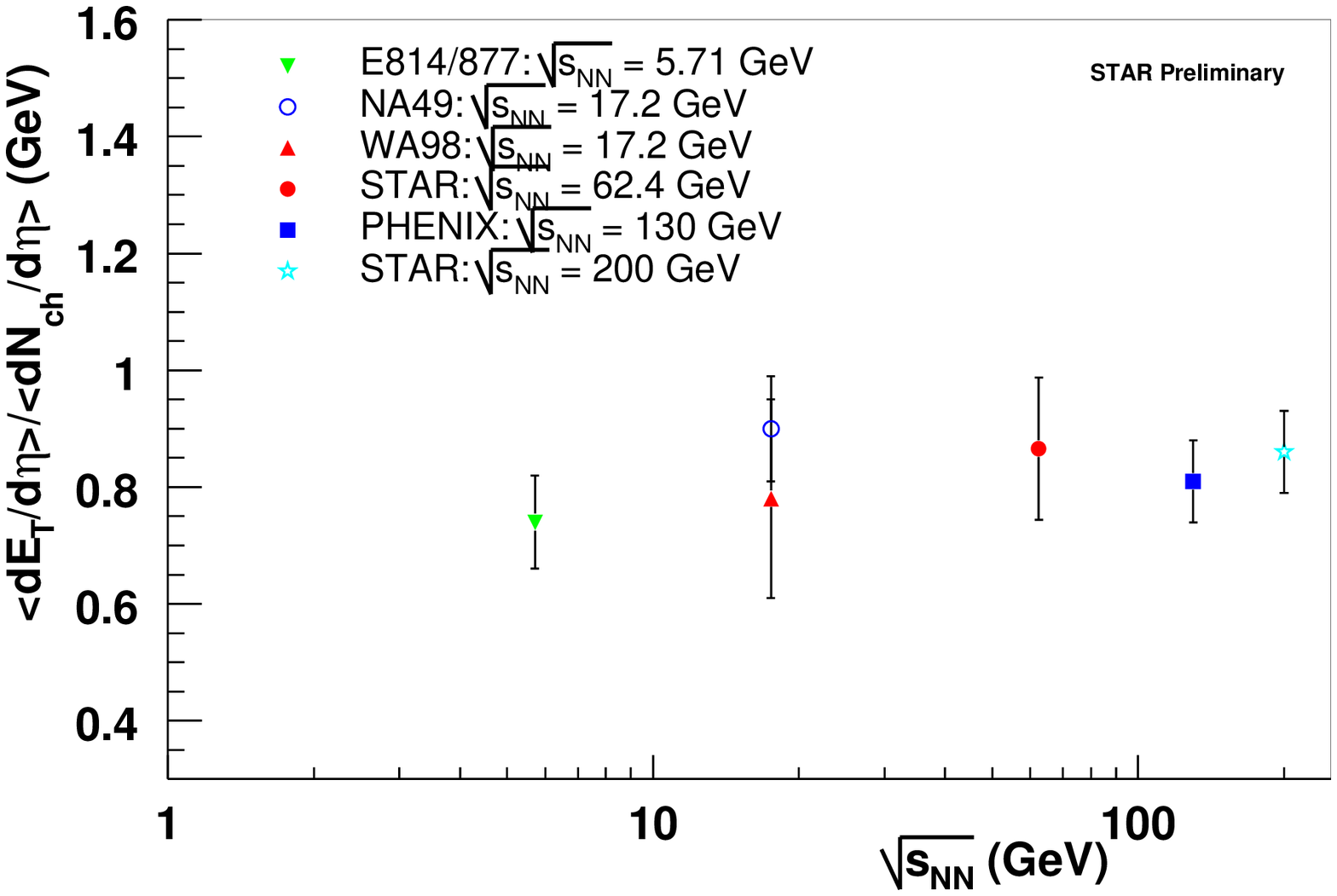}
\caption{$<dE_T/d\eta>/<dN_{ch}/d\eta>$ vs $N_{part}$ (left) and
$<dE_T/d\eta>/<dN_{ch}/d\eta>$ vs $\sqrt{s_{NN}}$.}
\label{etPerNch}
\end{center}
\end{figure}

In order to understand the systematic growth of transverse energy with $\sqrt{s_{NN}}$,
the centrality dependence of $<dE_T/d\eta>/<dN_{ch}/d\eta> \equiv E_T/N_{ch}$ is
shown in Figure~\ref{etPerNch} (left). $E_T/N_{ch}$ shows an centrality independent value
of $\sim 0.8$ GeV except for peripheral collisions which are affected by hydrodynamic
flow. The assumption of an isentropic expansion of the fireball explains the 
the decrease in $E_T/N_{ch}$ for peripheral collisions.
A similar centrality behavior of $<p_T>$ \cite{starEt} suggests that energy goes for 
new particle production, instead of increasing the mean energy of the particle. 
Taking top central events, the value of $E_T/N_{ch}$ for different energies is
shown in Figure~\ref{etPerNch} (right) as a function of $\sqrt{s_{NN}}$ from SPS to RHIC. 
This ratio is an important barometric measure of internal pressure in ultra-dense matter 
produced in heavy ion collisions \cite{gulassy}. This centrality and energy independence 
of $E_T/N_{ch}$ has been explained by statistical hadron gas model and is found to be 
associated with the freeze-out 
of the fireball \cite{myPLB,myEPJ}. The most important information from this observation 
is that, irrespective of the initial conditions which are controlled by the 
$\sqrt{s_{NN}}$ and centrality, the system evolves to the same final state at freeze-out. 
The observed saturated value of $E_T/N_{ch}$ could be taken as its value in the 
pre-hadronic state, as $T_{ch} \sim T_c$, the critical temeprature for the quark-hadron 
phase transition.

\begin{figure}
\begin{center}
\includegraphics[width=2.3in]{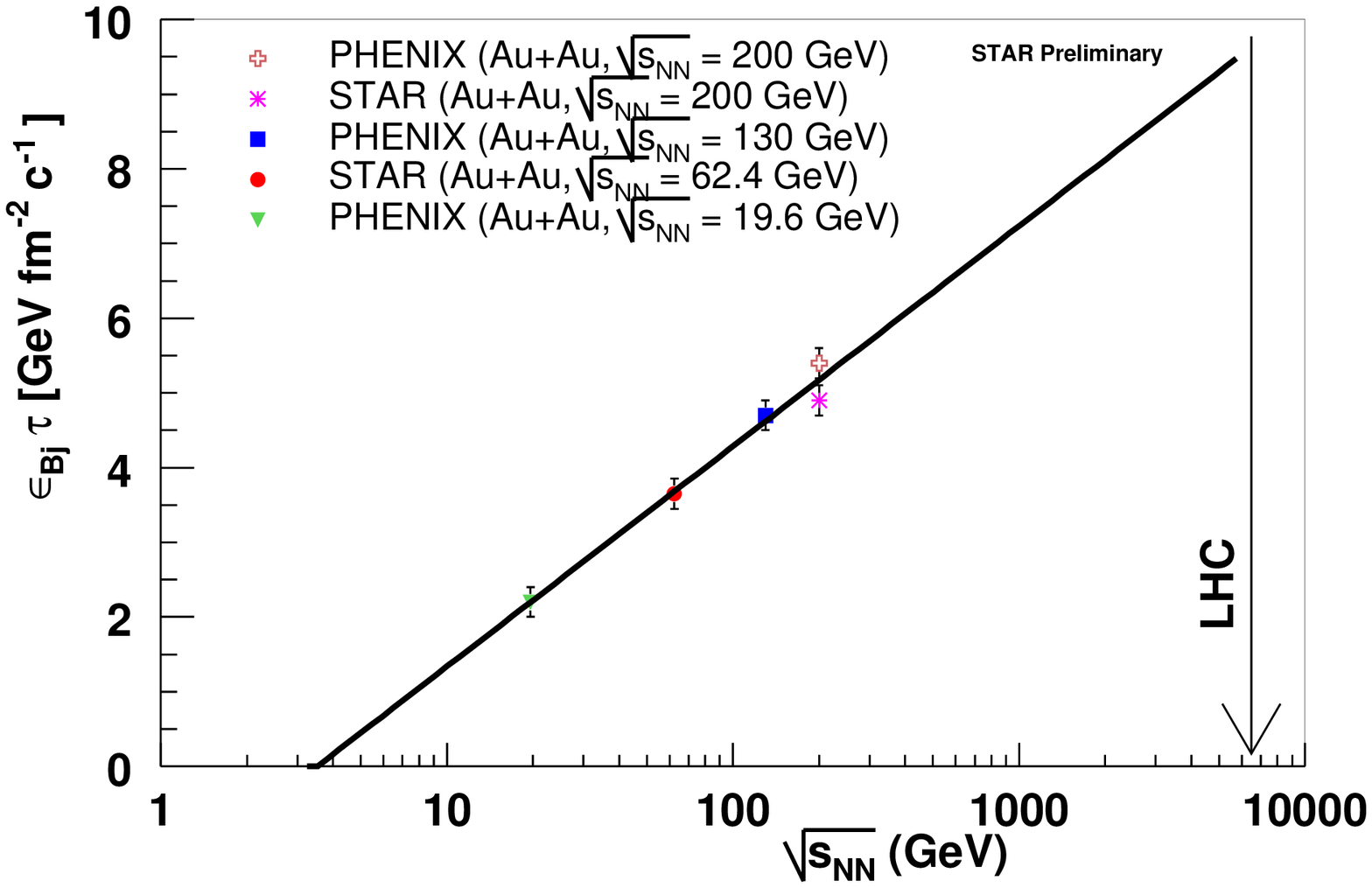}
\includegraphics[width=2.3in]{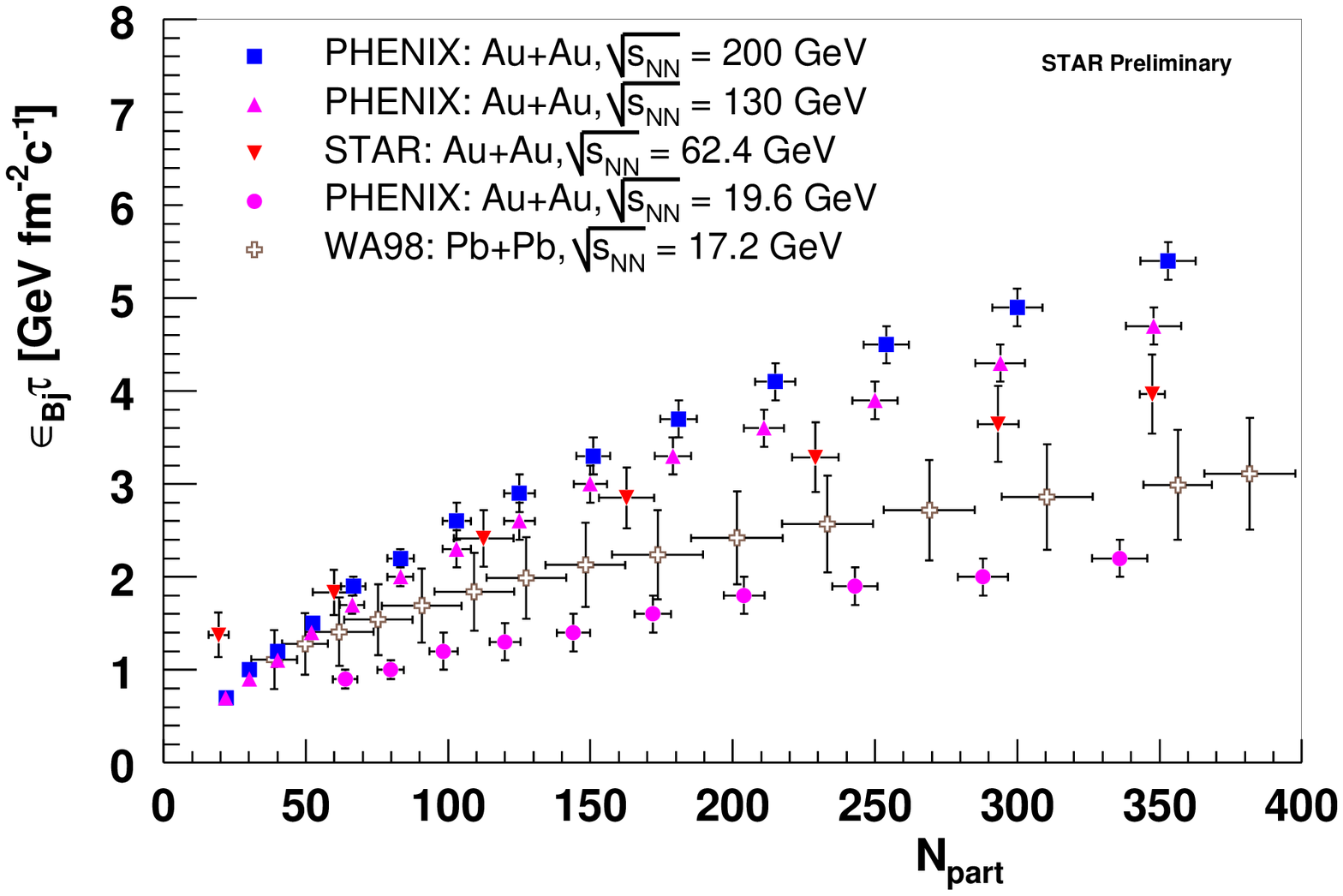}
\caption{$<dE_T^{em}/d\eta>/<dE_T/d\eta>$ vs $\sqrt{s_{NN}}$ for a number
of systems from SPS to RHIC (left) and $<dE_T^{em}/d\eta>/<dE_T/d\eta>$ vs 
$N_{part}$.}
\label{Bj}
\end{center}
\end{figure}

Next we have studied the Bjorken energy density \cite{jd} produced in the top central Au+Au 
collisions, which is shown in Figure~\ref{Bj} (left) along with similar data at other
RHIC energies. The value of the energy density produced in top central Au+Au collisions
at 62.4 GeV is found to be $3.65 \pm 0.39 ~GeV/fm^3$ (taking formation time, 
$\tau ~= ~1~ fm/c$). This is well above the lattice
QCD predictions \cite{lQCD} for a decofined state of quarks and gluons. This figure shows 
an logarithmic increase of $\epsilon_{Bj}.\tau$ and its extrapolation to LHC energy 
gives a value of $9.42\pm 0.55~GeV~fm^{-2}~c^{-1}$, based on the assumption that 
Bjoken model holds good at higher energies. We have shown the centrality dependence of
$\epsilon_{Bj}.\tau$ in Figure~\ref{Bj} (right), where one observes that higher energy 
densities are produced in more central collisions. Further more we 
have also studied the excitation functions of $<dE_T/d\eta>/(0.5N_{part})$ and 
$<dN_{ch}/d\eta>/(0.5N_{part})$ which show a logarithmic behavior. The details of which
could be found elsewhere \cite{thesis}. 

\begin{figure}
\begin{center}
\includegraphics[width=2.3in]{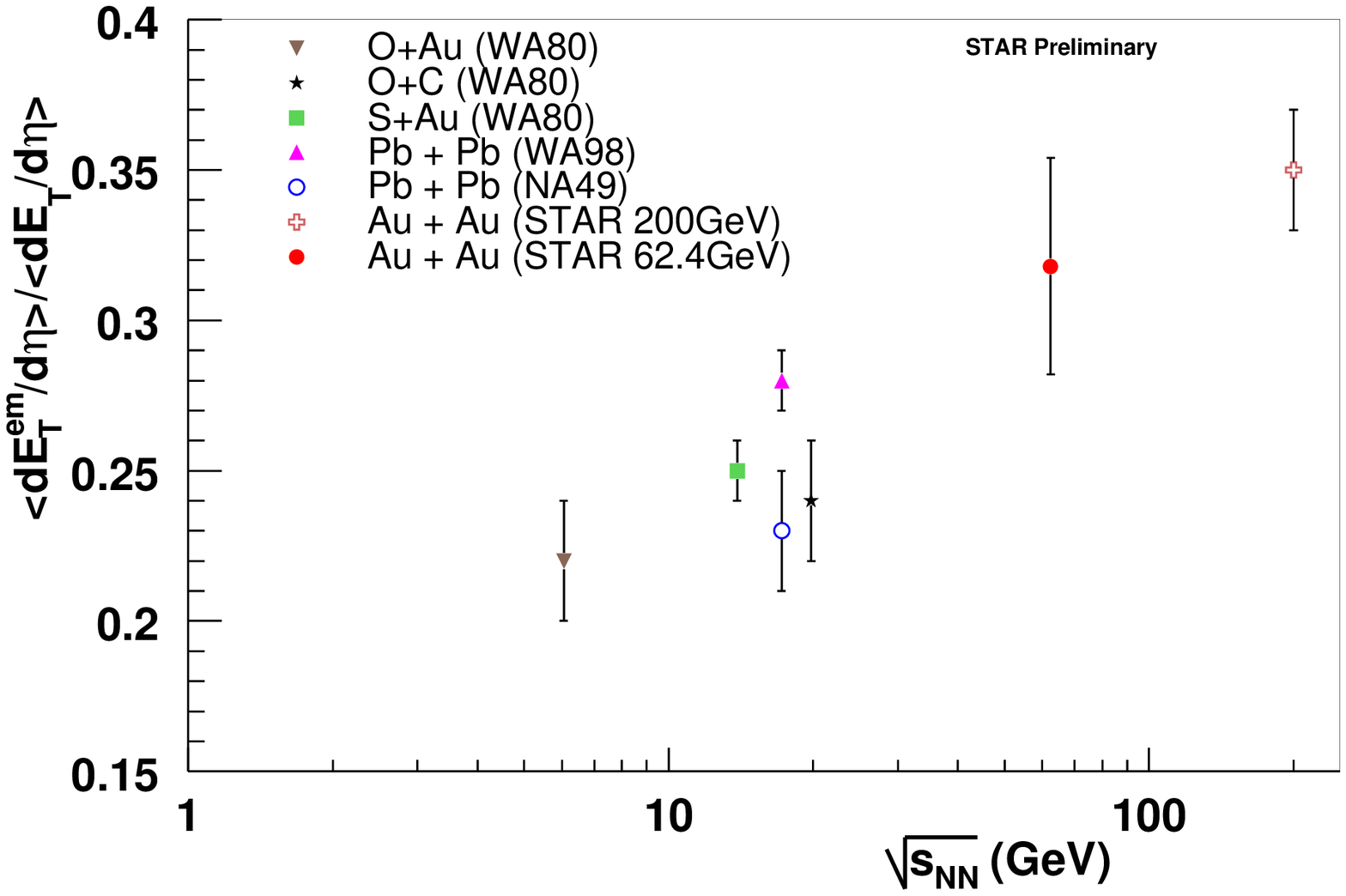}
\includegraphics[width=2.3in]{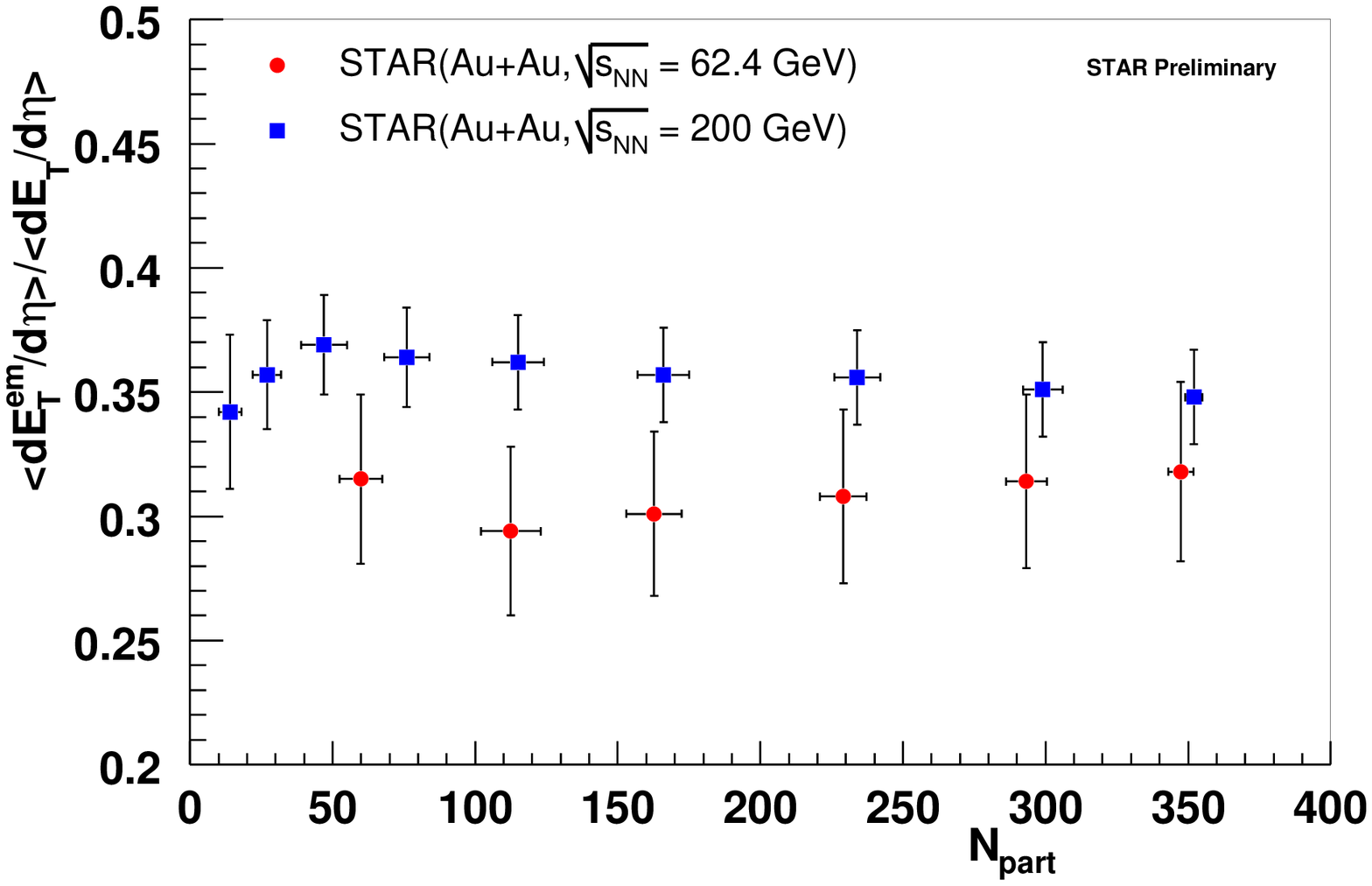}
\caption{$<dE_T^{em}/d\eta>/<dE_T/d\eta>$ vs $\sqrt{s_{NN}}$ for a number
of systems from SPS to RHIC (left) and $<dE_T^{em}/d\eta>/<dE_T/d\eta>$ vs 
$N_{part}$.}
\label{etEm}
\end{center}
\end{figure}

The electromagnetic fraction of the total transverse energy 
($<dE_T^{em}/d\eta>/<dE_T/d\eta>$) for top central Au+Au collisions for 62.4 GeV, is 
studied as a function of $\sqrt{s_{NN}}$, along with similar data from AGS-SPS to RHIC 
for different colliding species. This is shown in Figure~\ref{etEm} (left). The value 
of this ratio at 62.4 GeV is $0.32 \pm 0.03$ and the ratio shows a very slow increase with 
energy. This increase is consistent with the meson dominance of the matter at higher
energies. Figure~\ref{etEm} (right) shows that this ratio is independent of the 
collision centrality.
 
\section{Summary}
The mid-rapidity measurement of $E_T$ for 62.4 GeV Au+Au collisions is presented.
The centrality and center of mass energy behavior of $E_T$
production is studied and compared with similar data at other energies and
with the EKRT gluon saturation model. The observation of a centrality and $\sqrt{s_{NN}}$
independent, nearly constant value of $E_T/N_{ch} \sim 0.8$ GeV from AGS, SPS to RHIC has
been understood to be associated with freeze-out of the fireball. 
The initial energy density estimated by the boost-invariant Bjorken 
hydrodynamic model, is well above the lattice QCD value for a deconfined 
matter of quarks and gluons. Taking similar colliding species i.e. Au+Au, 
the $\epsilon_{Bj} .\tau$ has been predicted for LHC, based on the 
measurements at RHIC.\\

\end{document}